# An Empirical Comparison of the Summarization Power of Graph Clustering Methods

**Yike Liu**
University of Michigan
yikeliu@umich.edu

**Neil Shah**
Carnegie Mellon University
neilshah@cs.cmu.edu

**Danai Koutra**
University of Michigan
dkoutra@umich.edu

## Abstract

How do graph clustering techniques compare with respect to their summarization power? How well can they summarize a million-node graph with a few representative structures? Graph clustering or community detection algorithms can summarize a graph in terms of coherent and tightly connected clusters. In this paper, we compare and contrast different techniques: METIS, Louvain, spectral clustering, SlashBurn and KCBC, our proposed k-core-based clustering method. Unlike prior work that focuses on various measures of cluster quality, we use vocabulary structures that often appear in real graphs and the Minimum Description Length (MDL) principle to obtain a graph summary per clustering method.

Our main contributions are: (i) *Formulation*: We propose a summarization-based evaluation of clustering methods. Our method, VOG-OVERLAP, concisely summarizes graphs in terms of their important structures which lead to small edge overlap, and large node/edge coverage; (ii) *Algorithm*: we introduce KCBC, a graph decomposition technique, in the heart of which lies the k-core algorithm (iii) *Evaluation*: We compare the summarization power of five clustering techniques on large real graphs, and analyze their compression performance, summary statistics and runtimes.

## 1 Introduction

Summarization becomes increasingly crucial with the continuous generation of large amounts of data [15], as it can abstract away noise, and help discover existing patterns, which in turn may inform the human decision processes, as well as the design of new large-scale analysis algorithms. In this paper we focus on summarization of graphs, which are powerful structures that capture a host of phenomena, from communication between people to interactions between neurons in our brains [2, 13, 25]. Specifically, we put various graph clustering and community detection methods under the microscope and study their summarization power. Which graph clustering approach helps to summarize a graph in the best way? How expressive are the resulting summaries?

Detecting clusters or communities in graphs is of great interest in various domains, including social, biological, and web sciences [3, 11, 13]. There is very active research in the area, and numerous objective functions have been proposed to detect clusters or communities, that are defined as "tightly-connected" subgraphs. These clusters provide a better understanding of the underlying network structure (e.g. functional units in biology, research communities in collaboration networks), but can also be seen as a summary of the original graph. In this work we focus on the latter interpretation, and propose a novel way of comparing graph clustering methods. Unlike the literature which is rich in comparisons of clustering approaches in terms of cluster quality (e.g. modularity, average clustering coefficient, external and internal conductance, normalized mutual information), we compare and contrast the *output summaries* quantitatively and qualitatively. The idea is that the method that leads to the best summary is the one that helps the most in detecting and guiding a practicioner's attention to interesting and useful patterns.

For the graph summary generation, we leverage VoG [19, 20], a graph summarization algorithm that aims at succinctly describing million-node graphs. In a nutshell, VoG formulates the graph summarization problem as an information-theoretic optimization problem, where the goal is to find the hidden local structures that collectively minimize the global description length of the graph. For compression, VoG uses the Minimum Description Length (MDL) principle:

$$min \ L(G, M) = min\{L(M) + L(\mathbf{E})\}$$

where $\mathbf{M}$ is an approximation of $\mathbf{A}$ deduced by the model $M$, and $\mathbf{E} = \mathbf{M} \oplus \mathbf{A}$ is the error matrix. In addition to the MDL, the method uses a fixed vocabulary of structures $\Omega$ (cliques and near-cliques, stars, chains, and full or near-bipartite cores) that are ubiquitous in real-world graphs, and attempts to summarize the input graph in terms of these structures.

In [20], the first step of the summarization algorithm is to apply a subgraph extraction method, the output of which is then processed and refined by using MDL and a structure selection process. The authors adapted a node reordering method called SlashBurn [16], which biases the subgraphs to be stars or star-like – which is evidenced by the experiments in [20]. We propose to use VoG as a proxy to compare community detection and clustering methods with respect to a new metric, their summarization power: (i) SlashBurn and its adaptation to graph clustering [16, 20]; (ii) k-core-based clustering; (iii) Louvain community detection [4]; (iv) spectral clustering [14]; (v) METIS [17]. For k-core decomposition, we formulate it in a new way that provides multiple strongly connected components. We note that each method employs a different objective function and biases the shape of the discovered subgraphs accordingly. Most of the methods detect well-connected clusters that correspond to full cliques or bipartite cores. We propose using the MDL encoding cost as a quantitative means for comparing the methods: the most powerful method from a summarization perspective is the one that results in the best compressed graph summary. Our contributions are:

- *Formulation*: We are the first to evaluate clustering methods on the basis of MDL encoding, and thoroughly compare different graph decomposition techniques in terms of summarization power. We also introduce an edge-overlap-aware graph summarization method.
- *Algorithm*: We propose KCBC, a scalable, community detection algorithm based on $k$-cores.
- *Experiments*: We conduct thorough empirical analysis of five graph decomposition methods on large, real-world data.

The paper is organized as follows: In Section 2 we introduce the clustering methods that we consider, and our proposed k-core-based clustering algorithm. In Section 3 we introduce a new summarization model that accounts for overlapping edges in the extracted subgraphs. The empirical evaluation of the clustering methods is given in 4 followed by the related work and conclusions in Sections 5 and 6. We give the major symbols in Table 1.

**Table 1:** *Major symbols and definitions.*

| Notation | Description |
| --- | --- |
| $G(\mathcal{V}, \mathcal{E})$, $\mathbf{A}$ | graph, and its adjacency matrix |
| $\mathcal{V}, n = |\mathcal{V}|$ | node-set and number of nodes of $G$ resp. |
| $\mathcal{E}, m = |\mathcal{E}|$ | edge-set and number of edges of $G$ resp. |
| $\Omega$ | vocabulary of structure types |
| $M$ | a model for $G$ == a list of node sets with associated structure types |
| $L(G, M)$ | # of bits to describe model $M$, and $G$ using $M$ |
| $L(M), L(s)$ | # of bits to describe model $M$ and structure $s$ resp. |
| c | # of clusters |
| t | # of iterations |

## 2 Clustering Methods for Graph Summarization

One important constituent part of the summarization approach that we leverage, VoG, is the graph decomposition method that is used to obtain candidate structures for the graph summary. In this work, we study the effect of various clustering methods to the quality of the summary, and, in reverse, we use VoG-Contrast (Algorithm 1), a VoG-based approach, as a proxy to evaluate the summarization power of the clustering methods. We consider the following approaches: SlashBurn, out proposed k-core-based clustering (KCBC), Louvain, spectral clustering, and METIS. We first give a brief introduction of these algorithms and how we aim to leverage them to generate candidate subgraphs for the VoG-style summaries, and then summarize their qualitative comparison in Table 2.

**Algorithm 1** VoG-Contrast: Summarization-based Comparison of Graph Decomposition Methods

**Input**: graph $G$
**Step 1:** Summary extraction by using one of {SlashBurn, KCBC, Louvain, spectral clustering, and METIS}.
**Step 2:** MDL-based subgraph labeling by using a reduced vocabulary (compared to VoG[1]) that consists of full cliques, full bipartite cores, stars and chains.
**Step 3:** Summary assembly by employing the VoG heuristics Top10 and Greedy'nForget (Section 4).
**return** summary of the most important graph structures

**SlashBurn.** SlashBurn [16] is a node-reordering algorithm initially developed for the purpose of graph compression. The notion behind the algorithm is that removing the highest centrality nodes from the graph produces many small disconnected components and one giant connected component. SlashBurn performs two steps iteratively. First, it removes high centrality nodes from the graph. Next, it reorders nodes such that high-degree nodes are assigned the lowest IDs, nodes from disconnected components are assigned the highest IDs and nodes from the giant connected component are assigned the middle-range of IDs. In the next iteration, the process is repeated on the giant connected component. We leverage this process by identifying structures from the egonet[2] of each high centrality node, as well as the disconnected components as subgraphs.

**KCBC.** $k$-cores [12] have traditionally been used to unveil densely connected structures in graphs. A $k$-core can be defined as a maximal subgraph for which each node is connected to at least $k$ other nodes in the subgraph. Though the existence of $k$-cores in social graphs has been studied previously, to our knowledge no previous works have leveraged the $k$-core algorithm (recursively delete all nodes and adjacent edges in the graph of degree less than $k$) to identify communities in graphs. In this work, we develop a $k$-core-based algorithm to identify notable graph structures. The method is described in Algorithm 2. The main advantages of this method are that it is (i) fast and scalable with time complexity $O(n+m)$, (ii) can identify multiple structures per node and (iii) produces concise listings of non-redundant structures.

**Algorithm 2** KCBC: $k$-core-based Graph Clustering

**Input**: graph $G$
While the graph is nonempty
    **Step 1:** Compute core numbers (max $k$ for which the node is present in the decomposition) for all nodes in the graph.
    **Step 2:** Choose the maximum $k$ ($k_{max}$) for which the decomposition is non-empty, and identify nodes which are present in the decomposition as the "decomposition set." Terminate when $k_{max} = 1$.
    **Step 3:** For the induced subgraph from the decomposition set, identify each connected component as a structure.
    **Step 4:** Remove all edges in the graph between nodes in the decomposition set—they have been identified as structures already.
**return** set of all identified structures

**Louvain.** Louvain community detection [4] is a modularity-based graph partitioning method for detecting hierarchical community structure. The method is composed of two phases which are applied iteratively. In the first phase, each node is placed in its own community. Next, the neighbors $j$ of each node $i$ are considered, and $i$ is moved to $j$'s community if the move produces the maximum modularity gain. The process is applied repeatedly until no further gain is possible. In the second phase, a new graph is built whose supernodes represent the communities of the first phase, and superedges are weighted by the sum of weights of links between the two communities. The first phase is then applied to the new graph and the algorithm typically converges in a few such passes.

**Spectral Clustering.** Spectral clustering refers to a class of algorithms which utilize eigendecomposition of graphs or graph Laplacians to identify community structure. We utilize one such spectral clustering algorithm [14] which partitions a graph into $k$ segments by performs a $k$-means clustering on the top-$k$ eigenvectors of the input graph. The clustering is inspired by the notion that nodes with similar connectivity will have similar eigen-scores in the top-$k$ vectors and form clusters.

---

[1] Unlike VoG, in our algorithm we only consider full cliques, bipartite cores, stars, and chains. That is, we ignore the near-structures, because their MDL encoding incorporates error, which interferes with the global MDL encoding that we described in Section 1

[2] Egonet of a node is the induced subgraph of the node and its neighbors.

**METIS.** METIS [17] is a cut-based $k$-way multilevel graph partitioning scheme based on multilevel recursive bisection (MLRB) algorithms. Unlike other MLRB algorithms, METIS first coarsens the input graph by collapsing connected nodes into supernodes iteratively until the graph size is substantially reduced. Coarsening is done in a fashion to preserve edge-cut at each stage. Next, the coarsened graph is partitioned using MLRB and the partitioning is projected onto the original input graph $G$ by iteratively backtracking through the coarsened graph at each stage. In the uncoarsening phase, nodes are swapped between partitions to reduce edge-cut using the Kernighan-Lin algorithm. The method produces $k$ roughly equally sized partitions.

In Table 2 we compare the main features of these clustering methods, where $n$ is the number of nodes and $m$ is the number of edges in a graph. SLASHBURN and k-cores both give clusters that have overlapped edges, meaning in the model given by VOG where they are used as subgraph generation algorithms, there are likely to be structures including the same edges. Ideally we would like to minimize this kind of overlap between structures, which we explain and propose the encoding of such overlaps in description length.

**Table 2:** *Qualitative comparison of the clustering techniques.*

| Properties | Clustering Techniques | | | | |
|---|---|---|---|---|---|
| | **SlashBurn** | **KCBC** | **Louvain** | **Spectral Clustering** | **Metis** |
| **Overlapping Clusters** | ✔ | ✔ | ✗ | ✗ | ✗ |
| **Similar-sized Clusters** | ✗ | ✗ | ✗ | ✗ | ✗ |
| **Complexity** | $O(m + n \log n) \cdot t$ | $O(m + n)$ | $O(n \log n)$ | $O(n^3)$ | $O(m \cdot \# c)$ |
| **Parameter-free** | ✔ | ✔(in our alg.) | ✗ | ✔ | ✔ |
| **Number of Clusters** | High | Low | Medium | Medium | Medium |
| **Summarization Power** | Excellent | Poor | Good | Good | Good |
| **Cliques** | ✔ | ✔ | ✔ | ✔ | ✔ |
| **Bipartite Cores** | ✔ | ✗ | ✔ | ✔ | ✔ |
| **Stars** | ✔ | ✔ | ✔ | ✔ | ✔ |
| **Chains** | ✔ | ✗ | ✗ | ✗ | ✗ |

## 3 Encoding Overlapping Edges

In [20], the proposed summarization algorithm assumes that the candidate structures for the graph summary have node overlap, but no edge overlap. However, several graph decomposition methods, including SlashBurn and KCBC, produce edge-overlapping subgraphs, and there is good chance that a group of densely overlapping structures is chosen to be included in the graph summary because of they both appear to help with reducing the encoding cost of the graph. Moreover, the overlapping structures often result in lower coverage of nodes and edges than desired, and produce duplicated instead of comprehensive information for the input graph.

Our goal is to get a concise summary of the graph without explaining away edges multiple times – i.e. we want to minimize node and *edge* redundancy in our graph summary. We note that we are still interested in overlapping nodes that span multiple structures, as they can be seen as 'bridges' or 'connectors' and provide useful information about the network structure. To handle the above-mentioned issue, we propose a new method, VOG-OVERLAP, which extends VOG by minimizing the node/edge overlap and maximizing the coverage of the summary. First, we give an illustrative example that shows the issue of overlapping edges that arises from some graph clustering methods (in our case SlashBurn and KCBC), and then provide the details of VOG-OVERLAP, and show its better performance compared to VOG.

**An Illustrative Example.** Let us assume that the output of an edge-overlapping graph decomposition method is the following model which consists of three full cliques: *full clique 1* with nodes 1-20; *full clique 2* with nodes 11-30; and *full clique 3* with nodes 21-40. The VOG-based summary, which does not account for overlaps, includes in the summary all three structures, which clearly encode redundant nodes and edges. Despite the overlap, the model returns that it only needs 441 bits to describe the graph with the above-mentioned model, since it does not penalize edges that are covered multiple times. For reference, the graph needs 652 bits under the no model assumption. Ideally, we would want a method that penalizes extensive overlaps and tries to increase node/edge coverage.

**Encoding the Overlapping Edges.** To handle this issue, we propose VOG-OVERLAP which detects significant node and edge overlaps, and steers the structure selection process towards the desired output. We extend the optimization function for graph summarization by adding an overlap-related term (in bold):

$$min\ L(G, M) = min\{L(M) + L(\mathbf{E}) + \mathbf{L}(\mathbf{O})\}$$

where $\mathbf{M}$ is an approximation of $\mathbf{A}$ deduced by the model $M$, $\mathbf{E} = \mathbf{M} \oplus \mathbf{A}$ is the error matrix, and $\mathbf{O}$ is a weighted matrix that keeps track of the number of times each edge has been explain by $M$.

For consistency with VOG, we use the optimal prefix code [9] to encode the total number of overlapping edges. Following the literature [21], to encode the weights in matrix $\mathbf{O}$ which correspond to the number of times that each of the edges has been covered by the model, we use the MDL optimal encoding for integers [27]. The encoding for the overlaps is given by:

$$L(\mathbf{O}) = \log(|\mathbf{O}|) + ||\mathbf{O}||l_1 + ||\mathbf{O}||'l_0 + \sum_{o \in \mathcal{E}(\mathbf{O})} L_\mathbb{N}(|o|),$$

where $|\mathbf{O}|$ is the number of (distinct) overlapping edges, $||\mathbf{O}||$ and $||\mathbf{O}||'$ correspond to the number of present and missing edges in $\mathbf{O}$, $l_1 = -\log((||\mathbf{O}||/(||\mathbf{O}|| + ||\mathbf{O}||'))$, and analogue for $l_0$, are the lengths of the optimal prefix codes for the present and missing edges, respectively, and $\mathcal{E}(\mathbf{O})$ is the set of non-zero entries in matrix $\mathbf{O}$.

By applying VOG-OVERLAP to the example above, we obtain in the summary only the 1st and the 2nd clique, as desired. The encoding of our proposed method is 518 bits, which is higher than the number of bits of VOG: The reason is that in the VOG-OVERLAP-based summary some edges have remained unexplained (from nodes 11-20 to nodes 21-40), and, thus, are encoded as error. On the other hand, the VOG-based model encodes all nodes and edges (without errors), but explains many edges twice (e.g. the clique 11-20, the edges between 11-20 and 21-30) without 'counting' the redundancy-related bits twice.

## 4 Experiments

To compare and contrast the various decomposition methods used in VOG-CONTRAST, we use several real-world graphs which are presented with short descriptions in Table 3. We evaluate the clustering methods in three ways: (i) Trade-off between compression and coverage; (ii) Qualitative properties of their resulting summaries; and (iii) Runtime.

For structure selection we use the TOP10 and GREEDY'NFORGET heuristics that were introduced in [20]. Both heuristics order the candidate structures in decreasing order of encoding benefit (i.e. how much they help reduce the encoding cost of the graph). TOP10 returns the ten structures that help most, while GREEDY'NFORGET considers the structures sequentially and includes in the summary (or model $M$) only the ones that help further decrease the encoding cost of the graph. For all our experiments on Louvain, we choose resolution $\tau = 0.0001$.

Table 3: *Summary of graphs used in our empirical comparison.*

| Name | Nodes | Edges | Description |
| --- | --- | --- | --- |
| Flickr[3] | 404,733 | 2,110,078 | Friendship social network |
| Enron[3] | 80,163 | 288,364 | Enron email |
| AS-Oregon[3] | 13,579 | 37,448 | Router connections |
| Wikipedia-Chocolate | 2,899 | 5,467 | Co-editor graph |

### 4.1 Compression Rate vs. Node/Edge Coverage

We start by broaching the trade-off between compression rate and node/edge coverage for the five clustering methods that we consider. Table 4 gives the compression rate[4] of VOG-OVERLAP with the TOP10 selection heuristic, while Table 5 shows the compression rate of VOG-OVERLAP and the GREEDY'NFORGET selection. In the case of SlashBurn and KCBC, which result in edge-overlapping clusters, we provide the compression rate of VOG as well.

---

[3] http://www.flickr.com; http://www.cs.cmu.edu/∼enron; http://topology.eecs.umich.edu/data.html

[4] Compression rate refers to the ratio between the number of bits needed by the final model over the number of bits required to describe the graph under the empty model assumption.

Figure 1 shows the node and edge coverage of each VOG-OVERLAP + GREEDY'NFORGET summary model for the `Chocolate` and `AS-Oregon` graphs. The lighter and darker shades correspond to the node coverage before and after the structure selection (i.e. before step 3 of Algorithm 1). The node coverage for the non-overlapping clustering methods before the structure selection is not 100%, because we ignore clusters with fewer than 3 nodes, which are likely to be uninteresting from a practitioner's perspective.

**Observation 1** *In terms of compression rate, SLASHBURN and KCBC usually win over other methods, especially for the GREEDY'NFORGET heuristic. When KCBC has a much lower compression rate than SLASHBURN, it is due to the fact that KCBC covers very few nodes (e.g. nodes with degree*

**Table 4:** VOG-OVERLAP + TOP10: *Compression rate of the clustering techniques with respect to the empty model. The lower the rate, the better.*

| Dataset | Clustering Techniques | | | | |
|---|---|---|---|---|---|
| | **SLASHBURN** | **KCBC** | **Louvain** | **Spectral** | **METIS** |
| `Flickr` | **99%** | 100% | 100% | 100% | 100% |
| `Enron` | **98%** | 100% | 101% | 100% | 100% |
| `AS-Oregon` | **87%** | 96% | 100% | 100% | 104% |
| `Wikipedia-Chocolate` | 94% | **78%** | 101% | 101% | 104% |

**Table 5:** VOG-OVERLAP + GREEDY'NFORGET: *Compression rate of the clustering techniques with respect to the empty model. For KCBC, the rates in parentheses correspond to VOG encoding.*

| Dataset | Clustering Techniques | | | | |
|---|---|---|---|---|---|
| | **SLASHBURN** | **KCBC** | **Louvain** | **Spectral** | **METIS** |
| `Enron` | 75% (75%) | **41%** (39%) | 100% | 99% | 100% |
| `AS-Oregon` | 71% (71%) | **65%** (64%) | 95% | 94% | 96% |
| `Wikipedia-Chocolate` | 88% (88%) | **78%** (76%) | 99% | 99% | 100% |

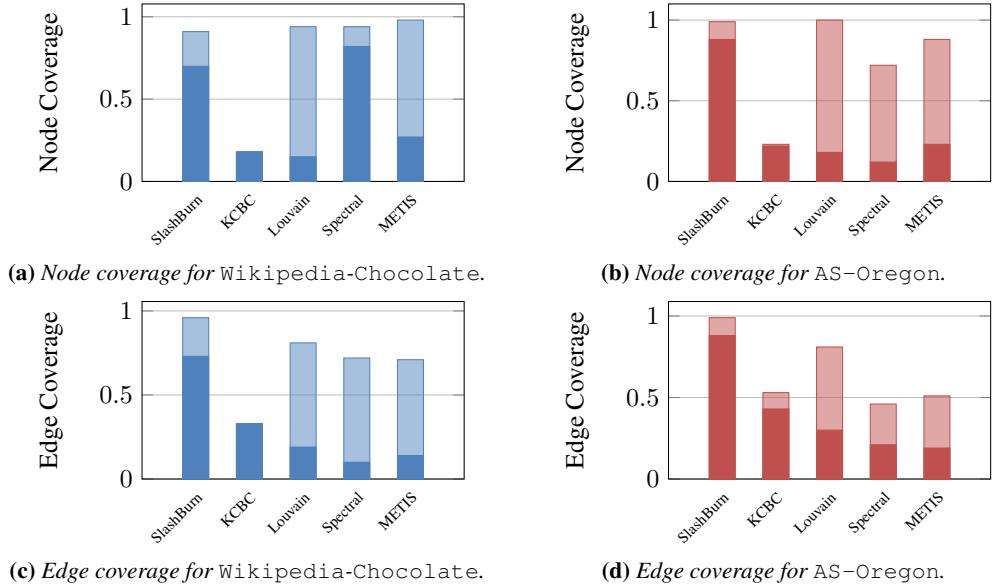

**(a)** *Node coverage for* `Wikipedia-Chocolate`.  **(b)** *Node coverage for* `AS-Oregon`.

**(c)** *Edge coverage for* `Wikipedia-Chocolate`.  **(d)** *Edge coverage for* `AS-Oregon`.

**Figure 1:** *SlashBurn results in summaries with the best node/edge coverage. Node and edge coverage for the five clustering methods and the* `Wikipedia`*-Wikipediachoc and* `AS-Oregon` *datasets.* *Light/dark color for before/after the structure selection.*

$\geq k$). *Though lower compression is usually better, we still want to get informative summaries that provide good coverage.*

We see this tradeoff clearly with SLASHBURN and KCBC. While KCBC has significantly lower compression rate, SLASHBURN achieves much higher node and edge coverage. Louvain and METIS lead to good coverage before the structure selection in step 3 of VOG-CONTRAST, but quite poor after that.

### 4.2 Qualitative Comparison of Structures

What types of structures do the summaries consist of? Which method is the most expressive in terms of summarization? Figures 2 and 3 depict the number of structures found by each clustering method before and after the selection step (light and dark color, respectively).

It is not surprising that SLASHBURN tends to find more structures than others, especially for stars, because of the way it decomposes the input graph. SLASHBURN contributes mainly stars and some cliques in the graph's final summary. KCBC's contribution is reverse. METIS, identifies both types of structures, but, as we will see in the next subsection, it is limited by its runtime.

**Observation 2** SLASHBURN *results in the most expressive summary which consists of all four types of structures. The othe approaches find mainly full cliques and bipartite cores (and before the selection step, they sometimes identify a few stars).*

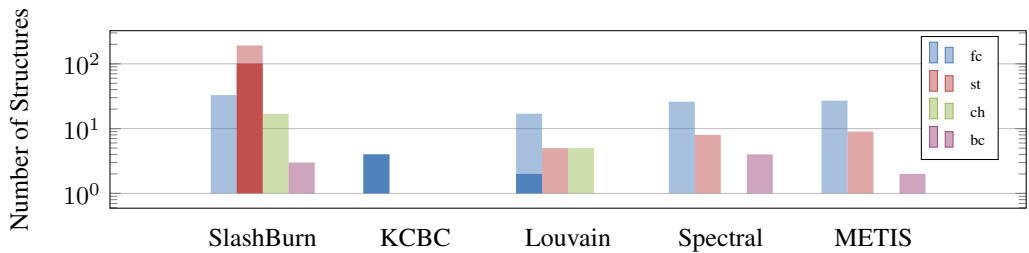

**Figure 2:** *Types of structures in the* `Chocolate` *graph by* VOG-CONTRAST. *Transparent/solid rectangles for before/after the structure selection step. Notation:: 'fc': full clique, 'st': star, 'ch': chain, 'bc': bipartite core.*

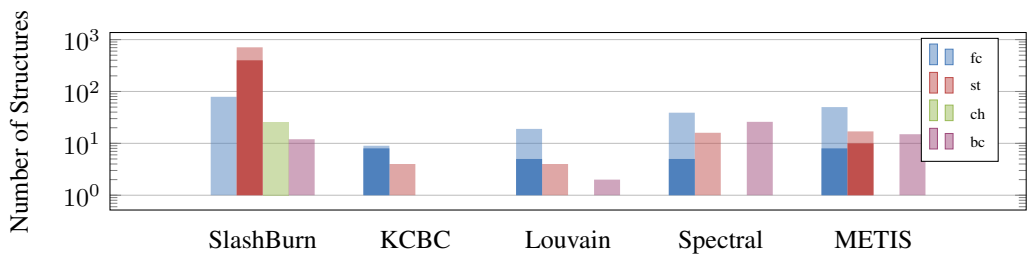

**Figure 3:** *Types of structures in the* `AS-Oregon` *graph by* VOG-CONTRAST. *Transparent/solid rectangles for before/after the structure selection step. Notation:: 'fc': full clique, 'st': star, 'ch': chain, 'bc': bipartite core.*

### 4.3 Runtime Comparison

In Table 6 we provide the runtime of the subgraph generation (clustering) and summary assembly for each method in VOG-CONTRAST. Since GREEDY'NFORGET tends to give better compression than the TOP10 heuristic, we only report result for that.

**Observation 3** *As far as the subgraph generation time is concerned, the ordering of the methods from the fastest methods are* KCBC*, spectral clustering and METIS, followed by Louvain and SlashBurn. The differences in the runtimes become bigger as the size of the input graph increases.*

*The summary assembly time is proportional to the number and size of candidate structures provided by the clustering method.*

Table 6: *Runtime of* VoG-Contrast +Greedy'nForget *for the five clustering techniques. We give the total runtime, and in parentheses the time division between structure generation/labeling (steps 1-2) and summary assembly (step 3). The fastest approaches are in bold.*

| Dataset | Clustering Techniques | | | | |
|---|---|---|---|---|---|
| | SlashBurn | KCBC | Louvain | Spectral Clustering | METIS |
| Enron | 81737.3s $(78108s + 3629.3s)$ | 11297.72s $(\mathbf{2859.8s} + 8437.92s)$ | 12771.5s $(5704.9s + 7066.6s)$ | **5067.08s** $(3661.1s + \mathbf{1405.98s})$ | 5443.95s $(5038.9s + \mathbf{405.05s})$ |
| AS-Oregon | 136.758s $(107.358s + 29.4s)$ | 36.3s $(\mathbf{6.71s} + 29.59s)$ | 237.9s $(8.22s + 229.68s)$ | 25.29s $(14.53s + 10.76s)$ | **24.08s** $(17.78s + \mathbf{6.30s})$ |
| Wikipedia -Chocolate | 5.8392s $(4.5092s + 1.33s)$ | 3.88s $(3.42s + \mathbf{0.46s})$ | 5s $(3.5s + 1.5s)$ | **1.9s** $(\mathbf{0.96s} + 0.94s)$ | 4.63s $(4.06s + 0.57s)$ |

**Observation 4** *Considering the summarization power generally,* SlashBurn *overpowers the other methods in quantity of discovered structures, especially stars. Cliques are identified by a selection of methods including* KCBC*, Louvain and spectral clustering, which are often faster than* SlashBurn*.*

## 5 Related Work

Work related to VoG-Contrast comprises MDL-based and graph clustering approaches.

**MDL and Graph Mining.** Many data mining problems are related to summarization and pattern discovery, and, thus, are intrinsically related to Kolmogorov complexity [10]. While not computable, it can be practically implemented by the Minimum Description Length principle [26] (lossless compression). Examples of applications in data mining include clustering [7], classification [22], community detection in matrices [6], and outlier detection [1].

**Graph Clustering.** We have already presented several graph clustering and community detection methods [4, 14, 17] in Section 2, which are all biased toward heavily connected subgraphs, such as cliques and bipartite cores. Other methods include the blockmodels representation [5], and community detection algorithms tailored to social, biological, and web networks [3, 11, 13]. Work on clustering attributed graphs [18, 28, 29] is related but does not apply in our case. Leskovec et al.'s work [23] is relevant to ours since it compares several clustering methods, but their focus is on classic measures of cluster quality, while we propose an evaluation of these method in terms of summarization power. Some summarization methods [8, 24] are related to VoG, but cannot be used as proxies to evaluate the summarization power of clustering techniques.

## 6 Conclusion

In this work we evaluate various graph clustering and community detection methods in terms of summarization power, in contrast to the literature that has focused on measures of cluster quality. We have proposed an MDL-based graph summarization approach, VoG-Contrast, which leverages the clusters found by graph decomposition methods, and is also edge-overlap aware (VoG-Overlap). Moreover, we have presented KCBC, a highly efficient, scalable and parameter-free graph clustering algorithm based on $k$-cores. Our thorough experimental analysis on real-world graphs has shown that each clustering approach has different strengths and weaknesses in terms of summarization power. Understanding their biases and combining them accordingly can give stronger better graph summaries with more diverse and high-quality structures.

## References

[1] L. Akoglu, H. Tong, J. Vreeken, and C. Faloutsos. Fast and Reliable Anomaly Detection in Categorical Data. In *CIKM*. ACM, 2012.

[2] L. Backstrom, D. P. Huttenlocher, J. M. Kleinberg, and X. Lan. Group formation in large social networks: membership, growth, and evolution. In *KDD*, pages 44–54, 2006.

[3] L. Backstrom, R. Kumar, C. Marlow, J. Novak, and A. Tomkins. Preferential behavior in online groups. In *WSDM '08: Proceedings of the international conference on Web search and web data mining*, pages 117–128, New York, NY, USA, 2008. ACM.

[4] V. D. Blondel, J.-L. Guillaume, R. Lambiotte, and E. Lefebvre. Fast Unfolding of Communities in Large Networks. *Journal of Statistical Mechanics: Theory and Experiment*, 2008(10):P10008, 2008.

[5] S. P. Borgatti. Special issue on blockmodels: Introduction. *Social Networks*, 14(1):1–3, 1992.

[6] D. Chakrabarti, S. Papadimitriou, D. S. Modha, and C. Faloutsos. Fully automatic cross-associations. In *Proceedings of the Tenth ACM SIGKDD International Conference on Knowledge Discovery and Data Mining, Seattle, Washington, USA, August 22-25, 2004*, pages 79–88, 2004.

[7] R. Cilibrasi and P. Vitányi. Clustering by Compression. *IEEE TIT*, 51(4):1523–1545, 2005.

[8] D. J. Cook and L. B. Holder. Substructure Discovery Using Minimum Description Length and Background Knowledge. *JAIR*, 1:231–255, 1994.

[9] T. M. Cover and J. A. Thomas. *Elements of information theory*. John Wiley & Sons, 2012.

[10] C. Faloutsos and V. Megalooikonomou. On Data Mining, Compression and Kolmogorov Complexity. In *Data Min. Knowl. Disc.*, volume 15, pages 3–20. Springer-Verlag, 2007.

[11] S. Fortunato. Community detection in graphs. *Physics Reports*, 486(3):75–174, 2010.

[12] C. Giatsidis, D. M. Thilikos, and M. Vazirgiannis. Evaluating cooperation in communities with the k-core structure. In *Advances in Social Networks Analysis and Mining (ASONAM), 2011 International Conference on*, pages 87–93. IEEE, 2011.

[13] M. Girvan and M. E. J. Newman. Community structure in social and biological networks. *PNAS*, 99:7821, 2002.

[14] J. P. Hespanha. An efficient matlab algorithm for graph partitioning. *Department of Electrical and Computer Engineering, University of California, Santa Barbara*, pages 25–67, 2004.

[15] H. V. Jagadish, J. Gehrke, A. Labrinidis, Y. Papakonstantinou, J. M. Patel, R. Ramakrishnan, and C. Shahabi. Big data and its technical challenges. *Commun. ACM*, 57(7):86–94, July 2014.

[16] U. Kang and C. Faloutsos. Beyond 'Caveman Communities': Hubs and Spokes for Graph Compression and Mining. In *ICDM*, 2011.

[17] G. Karypis and V. Kumar. Multilevel k-way Hypergraph Partitioning. pages 343–348, 1999.

[18] A. Koopman and A. Siebes. Characteristic Relational Patterns. In *KDD*, pages 437–446, 2009.

[19] D. Koutra, U. Kang, J. Vreeken, and C. Faloutsos. VoG: Summarizing and Understanding Large Graphs. In *SDM*, pages 91–99, 2014.

[20] D. Koutra, U. Kang, J. Vreeken, and C. Faloutsos. Summarizing and Understanding Large Graphs. In *Statistical Analysis and Data Mining*. John Wiley & Sons, Inc., 2015.

[21] E. Kuzey, J. Vreeken, and G. Weikum. A fresh look on knowledge bases: Distilling named events from news. In *Proceedings of the 23rd ACM International Conference on Conference on Information and Knowledge Management*, pages 1689–1698. ACM, 2014.

[22] M. van Leeuwen, J. Vreeken, and A. Siebes. Compression Picks the Item Sets that Matter. In *PKDD*, pages 585–592, 2006.

[23] J. Leskovec, K. J. Lang, and M. Mahoney. Empirical comparison of algorithms for network community detection. In *Proceedings of the 19th international conference on World wide web*, pages 631–640. ACM, 2010.

[24] S. Navlakha, R. Rastogi, and N. Shrivastava. Graph Summarization with Bounded Error. In *SIGMOD*, pages 419–432, 2008.

[25] OCP. Open Connectome Project. http://www.openconnectomeproject.org, 2014.

[26] J. Rissanen. Modeling by Shortest Data Description. *Automatica*, 14(1):465–471, 1978.

[27] J. Rissanen. A universal prior for integers and estimation by minimum description length. *The Annals of statistics*, pages 416–431, 1983.

[28] Z. Xu, Y. Ke, Y. Wang, H. Cheng, and J. Cheng. A model-based approach to attributed graph clustering. In *Proceedings of the 2012 ACM SIGMOD International Conference on Management of Data*, pages 505–516. ACM, 2012.

[29] Y. Zhou, H. Cheng, and J. X. Yu. Graph clustering based on structural/attribute similarities. *Proceedings of the VLDB Endowment*, 2(1):718–729, 2009.